# Long Term Friction: from Stick-Slip to Stable Sliding

Christophe Voisin[1], François Renard[1,2] and Jean-Robert Grasso[1]

1 LGIT-CNRS-OSUG, Université Joseph Fourier, Grenoble, France

2 Physics of Geological Processes, University of Oslo, Norway

Abstract. We have devised an original laboratory experiment where we investigate the frictional behaviour of a single crystal salt slider over a large number of deformation cycles. Because of its physical properties, salt, a surrogate for natural faults, allows for friction and plastic deformation and pressure solution creep to be efficient on the same timescale. During the same experiment, we observe a continuous change of the frictional behaviour of the slider under constant conditions of stiffness, temperature and loading velocity. The stick-slip regime is progressively vanishing, eventually reaching the stable sliding regime. Concomitantly, the contact interface, observed under the microscope, develops a striated morphology with contact asperities increase in length and width, arguing for an increase in the critical slip distance dc. Complementary experiments including velocity jumps show that the frictional parameters of the rate and state friction law, a and b, progressively vanish with the cumulative slip. In our experimental conditions, the ultimate stage of friction is therefore rate and state independent.

## 1. Introduction

Macroscopic solid friction obeys simple empirical laws, known as Amontons-Coulomb friction laws [*Amontons*, 1699; *Coulomb*, 1785]. They state the existence of a static threshold in friction and that friction depends on the normal load and not on the apparent contact surface area. Secondary effects have been reported since these laws were first proposed. At rest, the static friction coefficient $\mu_s$ increases with the logarithm of time [*Dieterich*, 1979]. This increase is contemporary to the plastic deformation of microscopic contact asperities under stress [*Dieterich and Kilgore*, 1994]. When sliding has begun, friction drops to a dynamic level $\mu_d$, which value is governed by the loading velocity. The most complete description of friction is encapsulated in the rate and state laws [*Dieterich*, 1979; *Rice*, 1983; *Ruina*, 1983]. These laws stipulate that friction depends on the slip velocity through two parameters *a* and *b*; and on a state variable $\theta$ that accounts for a mean-field description of memory effects of the interface:

$$\mu = \mu_0 + a\, ln\,(V/V_0) + b\, ln(V\,\theta/d_c),$$

where $\mu_0$ is a reference friction; $V_0$ is a reference velocity; V is the slider velocity; $d_c$ is a critical slip distance, akin to the mean size of asperities for instance.
Experimental studies have shown the existence of a stable and an unstable sliding regime, the latter known as the stick-slip mode. The observation of one or other of these modes is reported to depend on the stiffness of the experiment and on the driving velocity [*Dieterich*, 1979; *Heslot et al.*, 1994; *Marone*, 1998; *Shimamoto*, 1986]. As noticed by [*Shimamoto and Logan*, 1984], most of empirical friction laws are built on short-term experiments and their extrapolation to the geological time scale (long term) is highly speculative because some ductile processes (e.g. slow relaxation, pressure-solution, stress corrosion) are partially operative within the upper crust [e.g. *Gratier et al.*, 1999].

In the following, we present results from friction experiments, original in two ways: (i) we use a monocrystal of salt, both brittle and ductile at the laboratory timescale; (ii) the contact interface is observed under the microscope. We first describe the experimental apparatus. Second, we observe a continuous change from stick-slip to stable sliding as the slip cumulates under constant conditions of sliding velocity, normal load and temperature. Third, we observe that the microstructure of the contact interface evolves from randomly rough to some striated morphology. This process of ageing, not observed in plastic or elastic multicontact friction experiments, is driven in our experiments by salt pressure solution creep. Complementary friction experiments show that the frictional parameters *a* and *b* are decreasing as the slip cumulates. Finally, we propose a physical interpretation to the transition from unstable to stable slip as controlled by the physico-chemical ageing of the initially rough sliding surface.

## 2. Experimental Method

A cleaved monocrystal of halite (NaCl), roughened with sandpaper, is held under constant normal load and let in contact with a glass window (Figure 1). Salt is used because (i) it is transparent and allows for a direct observation of the contact interface; (ii) it behaves both in a brittle and a ductile way at low stress at ambient temperature and humidity [*Shimamoto*, 1986]; (iii) the plastic deformations are effective over the duration of the friction experiments. Using a salt slider allows for the brittle and ductile deformation to be effective on the time scale of our experiments, aimed as analogs of natural faults deforming in the brittle and ductile regimes.

The salt slider is mounted on an inverted microscope in order to observe the contact interface. Since halite is transparent, we image the contact asperities at the sliding surface using a high resolution camera located below the halite sample and focused at the slider interface undergoing shear. Doing so, we are able to track any visible deformation at the sliding surface. Horizontal and vertical displacements are monitored using linear encoder displacement transducers with sub-micron resolution. The shear resistance is recorded using a load cell (Figure 1). For all experiments the interface is subjected to ambient humidity. Under these conditions, a thin layer of water is adsorbed on the salt that promotes dissolution-crystallization reactions [*Foster and Ewing*, 2000]. During experiments the 1x1 $cm^2$ surface of the halite slider bore a constant normal load. Table 1 summarizes the experimental conditions of all runs. A first set of experiments is conducted at constant velocity. A second set of experiments imposes velocity jumps to the slider in order to estimate the frictional parameters *a* and *b*. All experiments are gouge-free, conducted with bare roughened salt sliders.

## 3. A Continuous Change from Stick-slip to Stable Sliding

### 3.1 Changes in the Frictional Behaviour of the Slider

Figure 2A plots the continuous variation in slip of a salt slider against glass, from stick-slip to stable sliding over several hundreds of deformation cycles. At the beginning of the experiment, the slider experiences regular stick-slip oscillations with 35 µm amplitude and 300 s waiting time (Figure 2B). The amplitude and waiting time gradually decrease as the slider enters the episodic stable sliding regime (Figure 2C, 2D and 3A). The waiting time decreases from 300 s to 60 s after 500 cycles, while the slip amplitude decreases from 35 µm to 10 µm. Most of this change occurs during the first 100 cycles. The process continues even when the stick-slip regime has disappeared and once the episodic stable sliding regime is established, with a decreasing period from 60 to 50 s and slip amplitude decreasing from 10 to 8 µm.

To investigate the robustness of this change of frictional behaviour, series of experiments are conducted with different imposed velocity, initial roughness, and different materials in contact (Figure 3B). All salt/glass experiments exhibit the same general trend of decreasing waiting time and slip amplitude, although the complete change towards stable sliding could not be observed for all experiments. Salt/PMMA experiments also follow the same change towards stable sliding, over a greater cumulated slip than for salt/glass experiments. PMMA being more compliant than glass, this suggests the importance of hardness contrast between the materials in contact to postpone the transition from stick-slip to stable sliding. On the contrary, PMMA/glass experiments have amplitude and waiting time that remain constant over the duration of the experiments. This last result is a positive evidence for the importance of the ductile processes to drive the change in frictional behaviour, much faster for salt than for PMMA under our laboratory experimental conditions.

**3.2 Changes in the Contact Interface Roughness**

The change from stick-slip to stable sliding is concomitant with the ageing of the contact interface. The roughness of the slider surface is measured before and after the experiment using white light interferometry (Figure 2A – colour insets). Initially, the surface root mean square (rms) of the roughness is close to 13.40 $\pm$ 0.05 µm. Contact asperities are separated by grooves caused by the roughening process. Their mean size in width and length is about 30 µm. This initial surface is representative of a multicontact interface [*Baumberger and Caroli*, 2006]. By the end of the experiment, for a cumulated slip of 0.6 cm, the sliding surface exhibits a roughness rms of 7.80 $\pm$ 0.05 µm. Contact asperities have grown and adopted an elongated shape in the direction of slip of dimensions 0.5 by 0.2 mm, giving the interface a striated morphology (Figure 2A – colour insets). The drastic change in surface morphology is accompanied by a downward vertical displacement of the slider. Inset in Figure 3A plots the power law relaxation of the vertical displacement with time. It is consistent with a deformation by pressure solution creep of the interface [*Dysthe et al.*, 2002]. The emergence of a strongly anisotropic morphology cannot be explained by the elasto-plastic ageing of the contact interface that leads to an isotropic growth of the contact asperities [*Berthoud et al.*, 1999; *Dieterich and Kilgore*, 1994]. The observed anisotropy arises from the coupling of pressure solution creep and horizontal displacement. Indeed, the change in topography is related to the development of the striated morphology of the contact interface. The matter dissolved from each contact area precipitates in the stress shadow of each asperity, leading to the observed anisotropic pattern.

**3.3 Changes in Frictional Parameters *a* and *b***

The two parameters *a* and *b* measure the velocity dependence of friction and the increase of static friction with hold time [*Dieterich*, 1979]. The difference (*b-a*) whenever positive implies a velocity weakening behavior, leading to stick-slip as observed at the beginning of our experiments. We conducted a series of experiments with velocity cycles (jumps from 1 to 10 µm/s) in order to measure these secondary effects of friction. Figure 4 plots three measures of *a* and *b*, for different cumulated slips. Both parameters are markedly decreasing with the cumulated slip. After a few centimeters of slip, the velocity jumps are hardly noticeable in the frictional behavior. During this final stage, the change in friction with velocity, if any, is insignificant. Therefore we cannot resolve whether the (*b-a*) difference changes sign. The slider has evolved from velocity weakening to velocity neutral.

## 4. Interpretation

Stick-slip and episodic stable sliding modes are described in the rate and state friction framework [*Dieterich*, 1979]. At constant driving velocity, the occurrence of one or other of these two frictional behaviors is related to a simple condition on the stiffness *K* of the experimental apparatus [*Heslot et al.*, 1994; *Scholz*, 2002]. To ensure stick-slip oscillations, one must have first *a* < *b*; second, *K* must obey the following relation:

$$K < K_c = W*(b-a)/d_c \qquad (1)$$

*W* stands for the normal load exerted on the slider, including its own mass. $d_c$ stands for a length scale typical of the interface, e. g. the mean size of asperities. Equation (1) arises from a stability analysis of a slider block under rate and state friction law [*Rice and Ruina*, 1983] and defines the value of the critical stiffness $K_c$ below which stick-slip oscillations do exist. Changes from stick-slip to episodic stable sliding are reported in various conditions. Changes in the driving velocity, in the stiffness or in the normal load all affect the frictional behavior [*Heslot et al.*, 1994]. The presence of a developed gouge also affects the occurrence of stick-slip or stable sliding by changing the (*b-a*) value [*Beeler et al.*, 1996; *Marone et al.*, 1990]. Temperature changes also affect the (*b-a*) difference [*Scholz*, 1998].

Our experiments are conducted under constant conditions of velocity, mass, stiffness and temperature. The low velocities we use preclude the wear of the slider and the development of a gouge. We must seek for other explanations to the observed transition from stick-slip to stable sliding. Since the normal load *W* is kept constant throughout the experiments, Equation (1) implies that $K_c$ has to decrease in order to explain the observed change from stick-slip to stable sliding. Our two observations, namely (i) the (*b-a*) decrease with time and (ii) the contact asperity size increase with time, both contribute to such a $K_c$ decrease with the cumulated slip. The direct observation of topography changes on the frictional interface (Figure 2A), akin to the growth of contact asperities, supports an increase in $d_c$ with time, that is a decrease in $K_c$. This increase in $d_c$ also implies a stabilization of the slider because of its finite size. Indeed, as $d_c$ increases the nucleation length increases as well, up to reach the stability limit [*Dascalu et al.*, 2000; *Voisin et al.*, 2002]. In the conditions of the experiment, the increase in $d_c$ is related to changes in the geometry of contact asperities, the latter being driven by pressure solution creep. This mechanism is solely responsible for a fast evolution of the interface under the low normal and shear stress conditions of our experiments [*Gratier*, 1993; *Karcz et al.*, 2006].

Although our experiments are performed under constant conditions of temperature and normal stress, and with no gouge, changes in (*b-a*) do occur. Indeed, the decrease in (*b-a*) is complete as both *a* and *b* vanish. The immediate consequence is that $K_c \rightarrow 0$ as the slip cumulates. Since (*b-a*)=$\partial\mu/\partial(\ln V)$, the ultimate frictional behavior of the salt slider is rate independent (Figure 4). A second consequence arises from the definition of *b*: $b=\partial\mu/\partial \log t$. Since $b \rightarrow 0$ as the slip cumulates, it implies that *μ* does not increase anymore at rest. The ultimate frictional behavior of the salt slider is state independent.

We explain the change from stick-slip to stable sliding as a progressive decrease in $K_c$ with the cumulative slip. Changing the stiffness of the experiment *K* will not preclude this evolution to occur. Indeed since $K_c \rightarrow 0$ with the cumulative slip, the change from stick-slip to stable sliding would be observed whatever the stiffness of the experiment. This hypothesis will be the concern of a future work.

## 5. Conclusion

The frictional behaviour of a single crystal salt slider is investigated under constant conditions of normal load, driving velocity, and temperature. We observe the progressive change from stick-slip to stable sliding with cumulative displacement. During the experiment, all frictional parameters are evolving: $a$ and $b$ are decreasing while $d_c$ is increasing. These changes are contemporary to the morphological evolution of the contact interface, i.e. the development a striated pattern driven by the coupling of pressure solution creep and slip. The increase in $d_c$ and the decrease in ($b$-$a$) both lead to the progressive vanishing of $K_c$. The salt slider is therefore forced to stable sliding, with no more rate and state dependence.

**Acknowledgments:** We thank P. Giroux, R. Guiguet, and L. Jenatton for their help in designing the experiment; D. Dysthe, J. Feder, D. Brito, J.-P. Gratier, L. Margerin, I. Manighetti , A. Helmstetter, K. Mair for thoughtful discussions. We acknowledge grants from CNRS (ATI, Dyeti) and from Université Joseph Fourier (BQR).

**Figure captions**

**Figure 1.** Schematic representation of the friction experiment. The 1 cm$^2$ surface area salt sample is housed in a plate made of a nickel iron alloy (Invar©) with a low thermal expansion coefficient to avoid thermal perturbations. The salt slider is carefully sawn from a single crystal of sodium chloride to avoid crack formation. The frictional interface is then roughened using different sandpapers (Struers #120, #320, #2400 grit) to ensure a variety of initial roughness. A constant continuous velocity (ranging from 0.11 up to 1.1 µm/s) is imposed on the plate through a brushless motor. The slider is subjected to a constant normal load (ranging from 0.11 to 0.24 MPa, user selectable) and is in contact with a glass or PMMA flat surface. Five displacement linear encoders record the plate movement in horizontal and vertical directions (LE/12/S IP50, Solartron). A force sensor (AEP TCA 5kg) records the shear force exerted to move the slider. The experiment is mounted on an inverted microscope (Olympus IX 70) equipped with a 12-bit CCD camera. The temperature is controlled as the whole set-up is let into a temperature-controlled Plexiglas box (21 $\pm$ 0.5°C). Humidity is ambient and not controlled.

**Figure 2:** (**A**) Change in frictional behavior of a salt/glass friction experiment. The constant imposed velocity is set to 0.11 µm.s$^{-1}$. The slider exhibits regular stick-slip at the beginning with a jump amplitude of about 35 µm and a waiting time of about 300 s. At the end experiment, the same slider exhibits an episodic stable sliding behavior with small oscillations of its speed around the imposed velocity. Color insets represent the topography of the frictional interface before and after the experiment measured by white light interferometry, with a roughness resolution of 0.05 µm (Wyko 2000 Surface Profiler from Veeco). Clear grooves created by the roughening procedure are visible. By the end of the experiment, the asperities have been flattened out and the average surface area of the contacts has increased. Note also the particular shape the contacts asperities have acquired during sliding. They are aligned parallel to the sliding direction. This striking change in the interface is related to the creep of asperities under normal and shear load. Column (**B**) stands for the beginning of the experiment. The cumulated slip is about 500 µm. The stick-slip behavior is clearly recorded both in horizontal and vertical displacements. Amplitudes are of the order of 20 µm and 0.3 µm in horizontal and vertical directions respectively. Force oscillations are also recorded with a period of 100 s. Long phases of stress build-up are followed by rapid force drops of up to 3 N as the slider moves abruptly. Column (**C**) stands for the mid-run of the experiment, with a cumulated slip of 2500 µm. The stick-slip behaviour is still recorded. Horizontal and vertical jumps are visible with amplitudes of about 10 µm and 0.1 µm respectively. The changes in force are quite similar to line (**C**) with rather smaller stress build-up and drops (less than 2 N) and a period of 80 s. Column (**D**) stands for the end of the experiment, with a cumulated displacement of about 6000 µm. Horizontal and vertical sensors record smooth oscillations typical of the episodic stable sliding regime with a period of 50 s. The force also exhibits smooth variations around 6.0 N. Note that the mean force that has to be exerted for the slider to move has increased: the friction coefficient has increased and the contact interface has strengthened. This is consistent with the increase of the real area of contact.

**Figure 3.** (**A**) Stick-slip amplitude ($\Delta d$) versus waiting time ($\Delta t$) for the friction experiment PUSH057. Color codes for increasing time: magenta; blue; red; yellow; and black. A clear trend to a decrease in amplitude and waiting time arises from the data. If there were no change process, all points would be located approximately at the same spot in this diagram. Since all experimental control parameters remain constant, the spreading of data along the observed trend characterizes the morphological change of the frictional interface. The red star (left lower corner) indicates the final state to be reached by the slider: At this stage, the slider moves at the exact imposed velocity. The inset shows the vertical displacement of the slider during the experiment. The overall decrease of the vertical displacement is consistent with the pressure solution creep of the interface. Second-order oscillations corresponding to stick-slip events also decrease with time.

**(B)** Stick-slip amplitude normalized by imposed velocity against waiting time. The trend towards stable sliding is observed in all experiments performed (Table 1), regardless of the nature of the materials in contact; salt/glass (circles) or salt/PMMA (squares), the imposed velocity $V$ (0.11 µm/s to 1.1 µm/s), the initial roughness of the slider defined by the sandpaper grit (#120 to #2400), and the normal load $W$ (1186-2651g). Note that PMMA/glass experiments (stars) do not exhibit the same evolution, which underlies the importance of the ductile deformation processes, much faster for salt than for PMMA in our experimental conditions.

**Figure 4.** Velocity jump experiments for $a$ and $b$ estimates. The slider is submitted to velocity cycles (1 and 10 µm/s). The black rectangles indicate the periods of slow velocity. The magnitude of friction change (Δµ) depends on the velocity jump and on the cumulative displacement. The three lines correspond to three exerts of an experiment (constant conditions of normal stress and temperature – no gouge development) taken at different times and cumulated displacement (Line 1: 0.25 to 0.45 cm; Line 2: 1.25 to 1.75 cm; Line 3: 4.2 to 4.7 cm). The rate of alternance is increased from Line 1 to Line 2. Sudden drops in friction observed at 1.53 and 4.5 cm corresponds to change in the direction of slip. Because of our experimental configuration, the driving force is reverted regularly to achieve a large cumulated slip. Frictional behaviour as well as its overall evolution is not affected when operating this way. For small cumulated displacement, the direct friction effect (related to $a$) and the slow relaxation (related to $b$) are visible. In Line 2, the direct friction effect has disappeared. The change in velocity induces an immediate change in friction that is also decreasing in amplitude with the cumulated displacement. Note that the slider is velocity weakening (µ=0.16 at 1 µm/s; µ=0.12 at 10 µm/s). In Line 3, the change in friction is no more visible. The slider has become velocity neutral.

**Table 1.** All experiments presented in this paper and shown in Figure 3B (PUSH experiments) and in Figure 4 (VJ experiments).

Figure 1

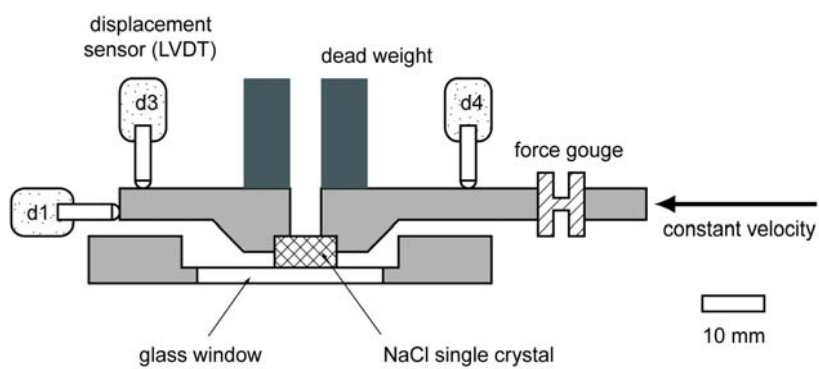

Figure 2

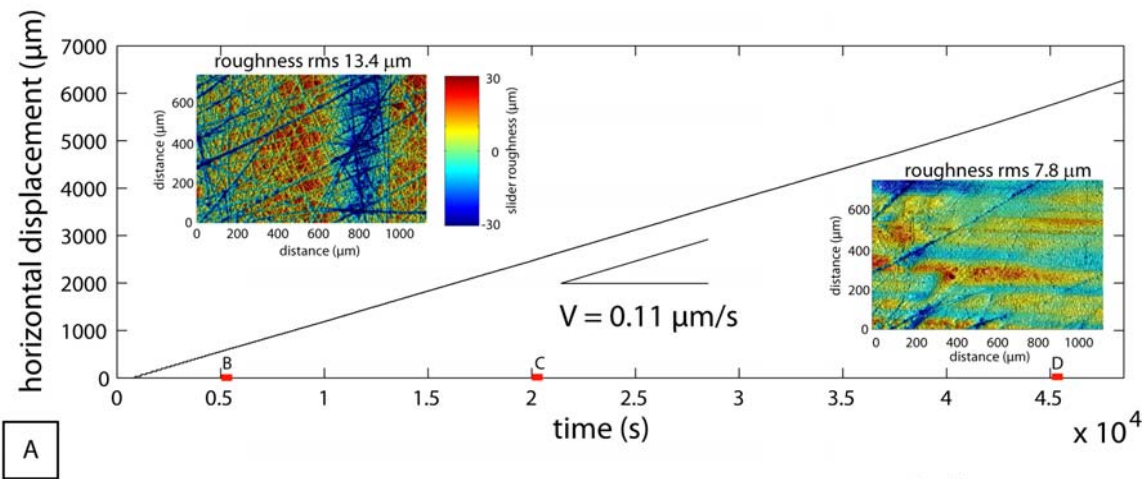

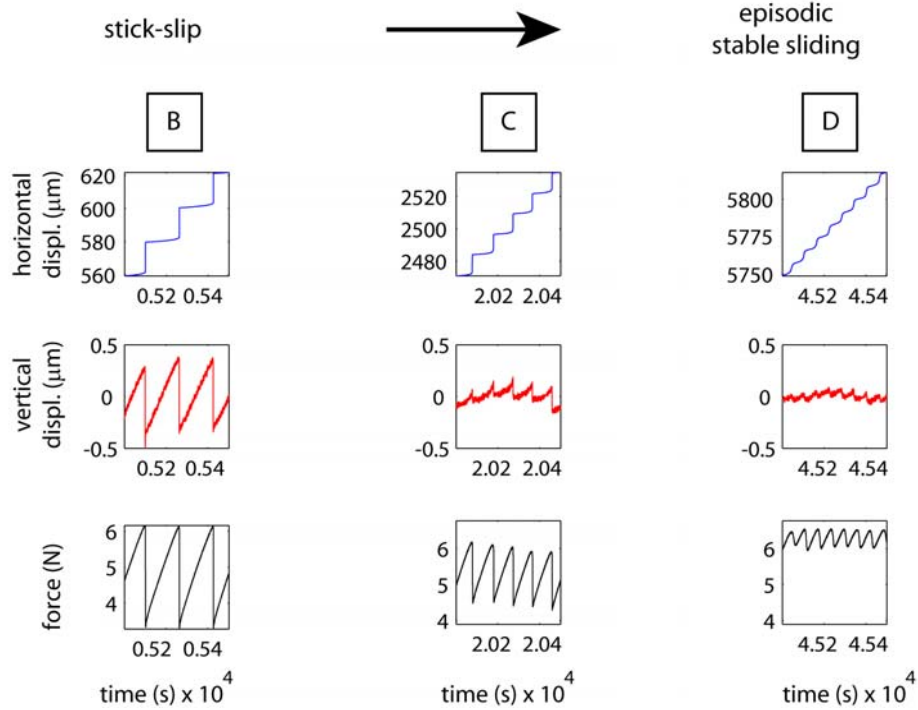

Figure 3A

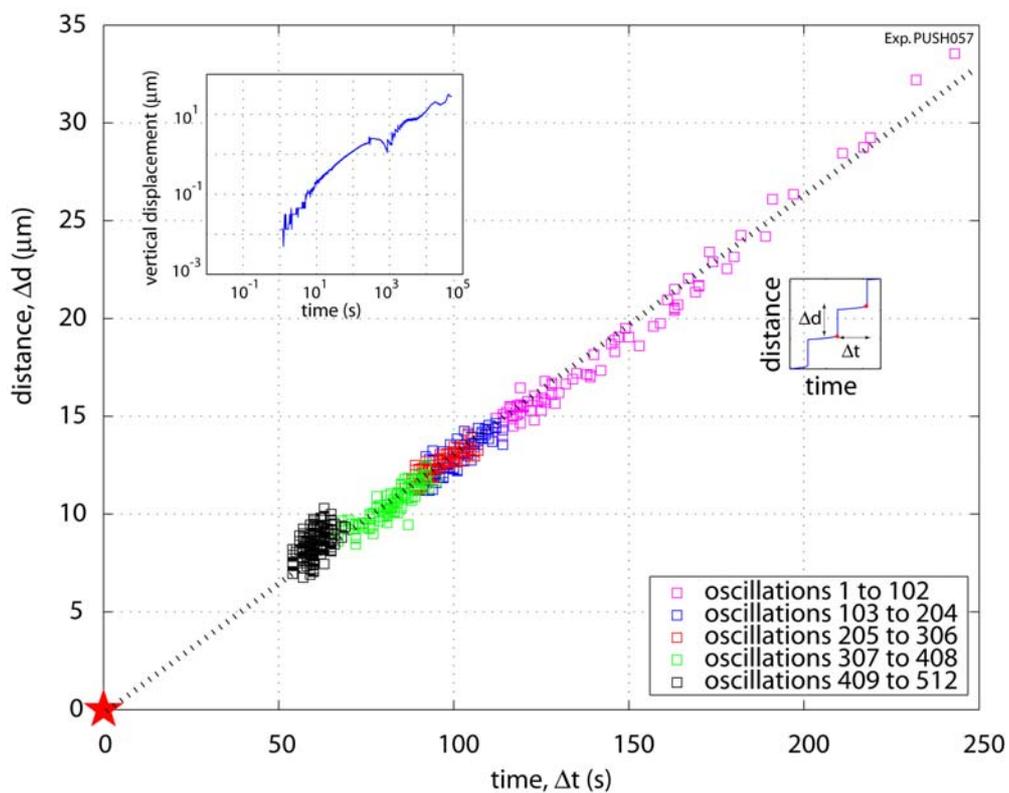

Figure 3B

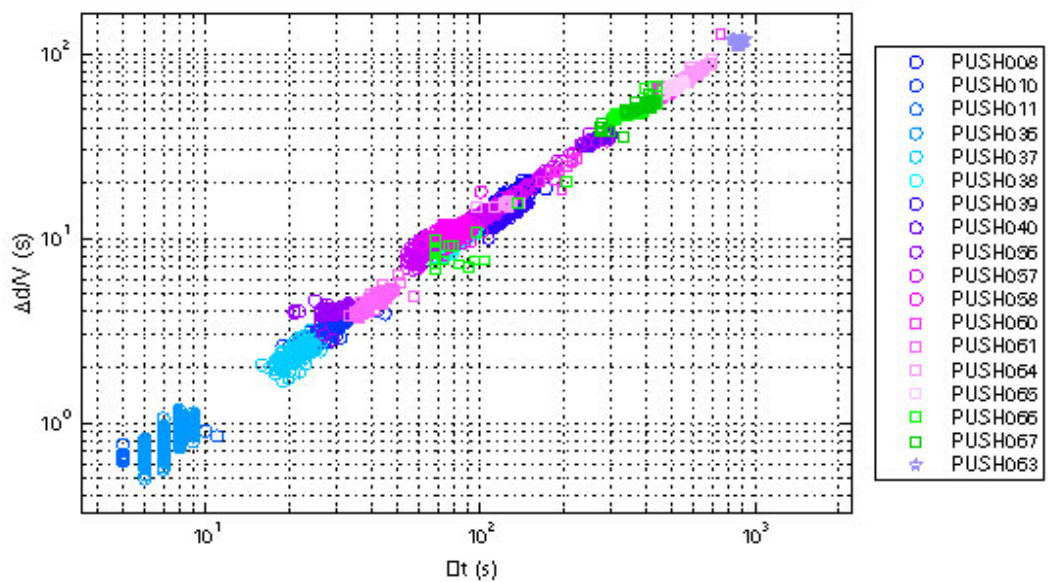

Figure 4

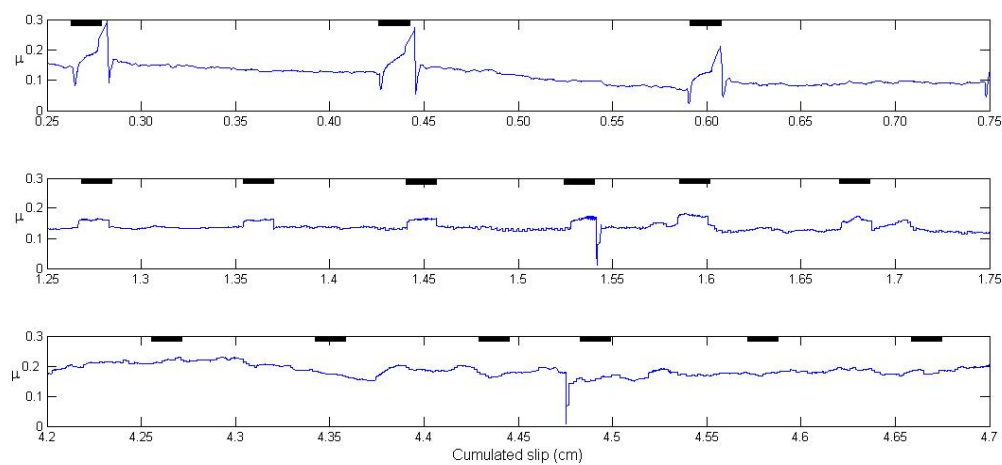

Table 1

| Experiment | Sample | Grit # | Plate | Weight (gram) | Motor velocity (µm/s) |
|---|---|---|---|---|---|
| PUSH008 | salt | 2400 | glass | 1186,1 | 0,11 |
| PUSH010 | salt | 2400 | glass | 1186,1 | 0,34 |
| PUSH011 | salt | 2400 | glass | 1186,1 | 1,13 |
| PUSH036 | salt | 2400 | glass | 2651,4 | 1,13 |
| PUSH037 | salt | 2400 | glass | 2651,4 | 0,56 |
| PUSH038 | salt | 2400 | glass | 2651,4 | 0,34 |
| PUSH039 | salt | 2400 | glass | 2651,4 | 0,22 |
| PUSH040 | salt | 2400 | glass | 2651,4 | 0,11 |
| PUSH056 | salt | 120 | glass | 2651,4 | 1,13 |
| PUSH057 | salt | 120 | glass | 2651,4 | 0,11 |
| PUSH058 | salt | 120 | glass | 2651,4 | 0,11 |
| PUSH060 | salt | 120 | PMMA | 2651,4 | 0,11 |
| PUSH061 | salt | 120 | PMMA | 2651,4 | 1,13 |
| PUSH063 | PMMA | 120 | PMMA | 2651,4 | 0,11 |
| PUSH064 | salt | 320 | PMMA | 2651,4 | 0,11 |
| PUSH065 | salt | 320 | PMMA | 2651,4 | 0,11 |
| PUSH066 | salt | 320 | PMMA | 2651,4 | 0,11 |
| PUSH067 | salt | 320 | PMMA | 2651,4 | 0,11 |
| VJ001a | salt | 320 | glass | 1186,1 | 1 -- 10 |
| VJ001b* | salt | 320 | glass | 1186,1 | 1 -- 10 |

* The rate of alternance is increased by a factor of 2. The same sample is used in VJ001a and VJ001b.